\documentclass[preprint,journal]{vgtc}       
\ieeedoi{10.1109/TVCG.2019.2934669}

\ifpdf
  \pdfoutput=1\relax                   
  \usepackage{graphicx}                
  \DeclareGraphicsExtensions{.pdf,.png,.jpg,.jpeg} 
\else
  \usepackage{graphicx}                
  \DeclareGraphicsExtensions{.eps}     
\fi%

\graphicspath{{figures/}{pictures/}{images/}{../}} 

\usepackage{microtype}                 
\PassOptionsToPackage{warn}{textcomp}  
\usepackage{textcomp}                  
\usepackage{mathptmx}                  
\usepackage{times}                     
\usepackage{cite}                      
\usepackage{tabu}                      
\usepackage{booktabs}                  
\usepackage[usenames,dvipsnames,svgnames,table,cmyk]{xcolor} 

\usepackage{tikz}
\usepackage[11pt]{moresize}      
\usepackage{hyperref}

\usepackage{tabularx}
\usepackage{booktabs}
\usepackage{array}
\usepackage{multirow}
\usepackage{colortbl}

\newcolumntype{L}{>{\raggedright\arraybackslash}X}
\newcolumntype{C}{>{\centering\arraybackslash}X}
\newcolumntype{R}{>{\raggedleft\arraybackslash}X} 
\usepackage{xcolor}
\usepackage{soul}
\definecolor{highlightColor}{rgb}{1, 0.901, 0.556}
\sethlcolor{highlightColor}

\newcommand{\inlinegraphic}[1]{%
	\protect\raisebox{-1.6pt}{%
		\protect\includegraphics[height=0.28cm]{#1}%
	}\,%
}

\usepackage[final]{changes}

\onlineid{0}

\vgtccategory{Research}
\vgtcpapertype{please specify}

\title{Exploranative Code Quality Documents}

\author{Haris Mumtaz, Shahid Latif, Fabian Beck, and Daniel Weiskopf}
\authorfooter{
\item
 H. Mumtaz and D. Weiskopf are with VISUS, University of Stuttgart, Germany. E-mail: \{haris.mumtaz, daniel.weiskopf\}@visus.uni-stuttgart.de.
\item
 S. Latif and F. Beck are with paluno, University of Duisburg-Essen, Germany. E-mail: \{shahid.latif, fabian.beck\}@paluno.uni-due.de.
}

\abstract{\replaced{Good code quality is a prerequisite for efficiently developing maintainable software.}{Information about the quality of source code is necessary to reduce the costs of maintenance and improve reusability in software engineering. Although there exist many code inspection and software quality tools, information is often presented in a way that lacks putting the data into context.} In this paper, we present a novel approach to generate \emph{exploranative} (explanatory and exploratory) data-driven documents that report code quality in an interactive, exploratory environment. We employ a template-based natural language generation method to create textual explanations about the code quality, dependent on data from software metrics. The interactive document is enriched by different kinds of visualization, including parallel coordinates plots and scatterplots for data exploration and graphics embedded into text. We devise an interaction model that allows users to explore code quality with consistent linking between text and visualizations; through integrated explanatory text, users are taught background knowledge about code quality aspects. Our approach to interactive documents was developed in a design study process that included software engineering and visual analytics experts. Although the solution is specific to the software engineering scenario, we discuss how the concept could generalize to multivariate data and report lessons learned in a broader scope.  %
} 


\CCScatlist{ 
 \CCScat{K.6.1}{Management of Computing and Information Systems}%
{Project and People Management}{Life Cycle};
 \CCScat{K.7.m}{The Computing Profession}{Miscellaneous}{Ethics}
}

\teaser{
  \centering
  \includegraphics[width=\linewidth, trim=0.0in 0.0in 0.0in 0in]{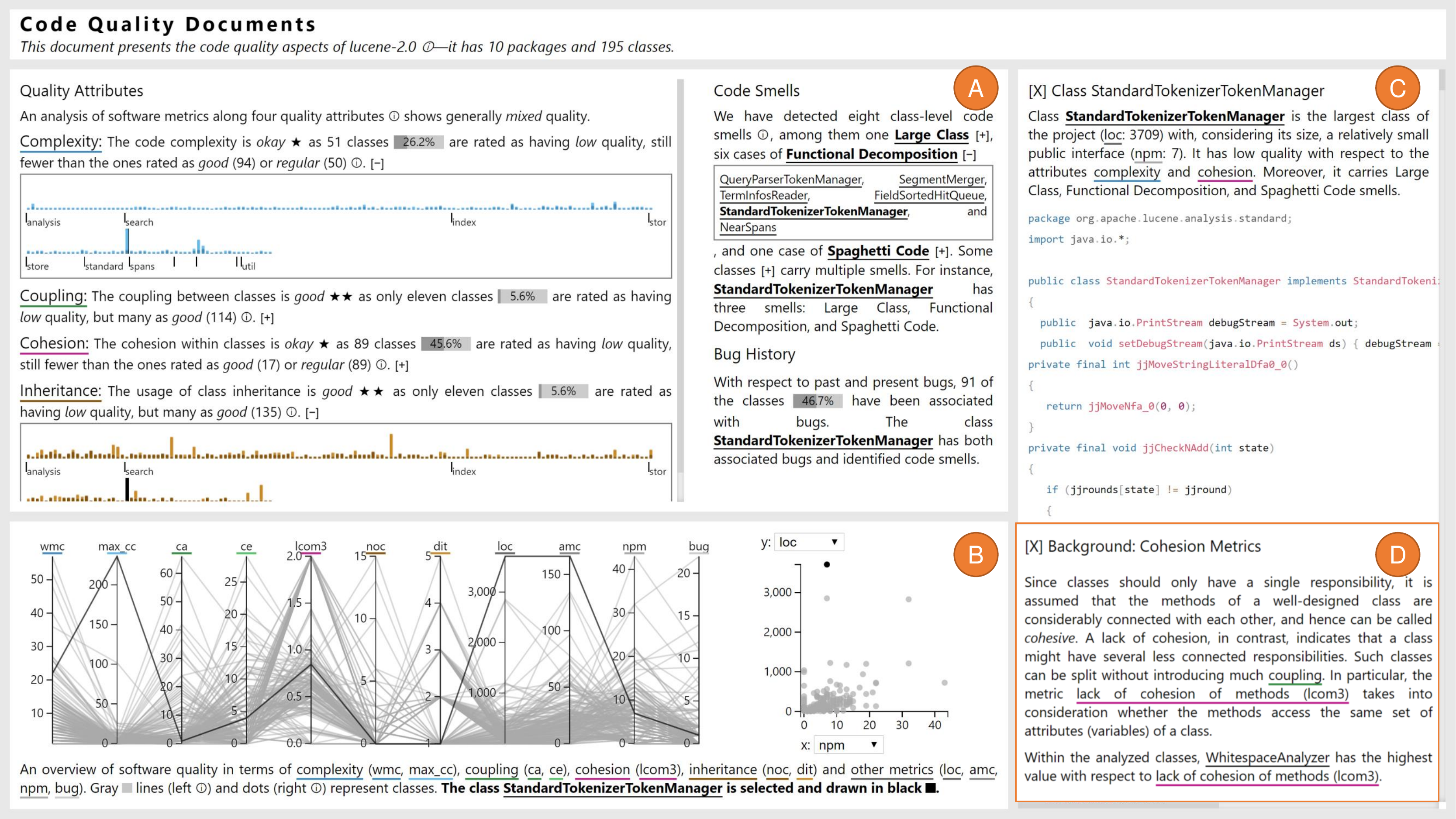}
  \caption{Exploranative code quality document for \emph{Lucene 2.0}. {\large\textcircled{\small A}}~Textual overview in terms of quality attributes, code smells, and bugs, which includes embedded visualizations. {\large\textcircled{\small B}}~Overview visualizations: parallel coordinates plot and scatterplot. {\large\textcircled{\small C}}~Source code of a class provided in the details view. {\large\textcircled{\small D}}~Description of a quality attribute alternatively presented in the details view.}
	\label{fig:teaser}
}

\begin{document}


\firstsection{Introduction} \label{sec:intro}
\maketitle
To create high-quality software, the stakeholders involved need to obtain information about code quality\deleted{ so that necessary measures can be taken to improve the parts of the source code that exhibit low quality}. 
\replaced{Existing visualization approaches provide an overview of metrics related to code quality, but they often lack to put data into context. Findings related to different code quality aspects need to get connected, explanations might be required for stakeholders less experienced in software quality, and making transparent the workings of analysis algorithms will increase trust into the quality assessment. In contrast, some tools integrated into the software development chain already make a step in this direction, however, consider different quality aspects separately and use only simple visualizations. Our goal is to go a step ahead and integrate visualizations with contextual information.}{Existing tools and visualization approaches already provide source code information related to bugs, test coverage, code smells, etc., but they lack to put data into context. For instance, tools may provide a list of issues in a system, but they do not highlight the importance of certain issues, connect them with other findings, or explain background information. However, contextual information is important to fully understand and highlight the important aspects of code quality. This leads to the questions of how such context can be presented to stakeholders. Existing solutions often provide information about the systems through visualizations, textual descriptions are minimal. In contrast,} We argue that textual descriptions are key to communicate context and guide the data exploration process.\added{ Additional textual explanations can make the interface fully self-explanatory, allow the integration of abstract information that is hard to explain visually, and facilitate blending domain terminology with data descriptions.}

We suggest a visual analytics solution that combines text and visualization. The idea is to automatically generate textual descriptions that guide the users through the main findings of a code quality analysis. We augment the generated text with visualizations that also allow users to interactively explore different aspects of source code quality. 
With this, we adopt the ideas of \emph{exploranation} (explanation and exploration)~\cite{ynnerman2018exploranation,Weiskopf:2006:relativity} and \emph{explorable explanations}~\cite{victor2011explorable}\deleted{, both originally focusing on education scenarios,} for a visual analytics application in a professional context. We describe our underlying concept as a general approach to represent multivariate data before we tailor it in a design study process to the specific application of code quality analysis.

In the visual analytics system shown in Figure~\ref{fig:teaser}, we use a multi-view interface to present different perspectives on the data. The overview panels on the left (Figure~\ref{fig:teaser} A and B) show code quality information in the form of text and visualizations. \added{The main text (Figure~\ref{fig:teaser} A) summarizes code quality attributes based on four important indicators (left) and discusses code smells and the bug history.} We use visualizations embedded in the text (A: small icons and bars in line with text and bar charts between the text lines) as well as traditional visual representations of multivariate data (B: a parallel coordinates plot and a scatterplot) \added{to report metric values for the classes of an object-oriented system. While the bar charts show a subset of related metrics per class (grouped by packages) in stacked bars, the parallel coordinates plot displays all metrics for all classes (i.e., each line represents a class); the scatterplot adds a different perspective for a user-selected pair of metrics. All visualizations are interactive and linked with each other to support further exploration. }Details are provided on demand in the right panel  (Figure~\ref{fig:teaser} C and D). For instance, the user can learn about the specifics of a class (C: textual description and source code) or get background information about a specific code smell (D: explanation and example). The navigation between the views is supported by multi-directional \emph{vis--text} interactions. \added{This means that the text is interactively linked with the visualization and vice versa; for example, hovering a class in the text or in any visualization highlights the other representatives of the same class across all views.} 
With these features, the system guides through the result of the code analysis and supports active exploration of the data. 
Overall, our paper provides the following main contributions:
\begin{itemize}
  \item A general concept building on textual explanations, explorable visualizations, and consistent linking between them to represent multivariate data as interactive documents.
  \item A visual analytics system that allows users to learn and understand code quality aspects in an interactive environment driven by textual explanations and visual exploration.
  \item An interaction model that provides close integration of text and visualization by interactive linking.
  \item Lessons learned from our design study that could impact the design of data-driven interactive documents.
\end{itemize}

We have followed a design process involving four authors with a diverse background in software engineering, visual analytics, interactive documents, and text generation. 
We have also conducted a formative study with different external experts to evaluate and improve our system in an iterative manner. 
Our system is available as a web application at \url{https://vis-tools.paluno.uni-due.de/cqd}. We also provide the system as well as the material and data of our formative study as supplemental material\cite{dataAndTool}.


\section{Related Work} \label{sec:rw}


Our overview of related work covers aspects of software engineering, visualization, and visual analytics, as well as interactive documents.
\deleted{ We introduce software metrics used for code quality analysis and discuss limitations of existing visualization solutions and tool interfaces for such data. With respect to interactive documents, we report on text generation, with a specific focus on approaches that are applied in the context of software engineering, combine generated text and visualization, or discuss interactions between text and visualization.}

\added{\textbf{Code Quality} --} Code quality is multi-faceted and covers, for instance, testability, maintainability, and readability. 
To examine these aspects, certain quality attributes (e.g., coupling, complexity, size) are quantified by underlying software metrics; for instance, McCabe's software complexity metrics measure readability aspects of the code~\cite{mccabe1976complexity}. For object-oriented systems, a popular set of metrics is the CK suite introduced by Chidamber and Kemerer~\cite{chidamber1994metrics} and the QMOOD metrics (Quality Model for Object-Oriented Design)~\cite{bansiya2002hierarchical}. Many approaches employ such metrics suites to distinguish parts of the source code in terms of good, acceptable, or bad quality~\cite{filo2015catalogue, shatnawi2010finding} or to identify code smells (problematic properties and anti-patterns of the code)~\cite{ouni2013maintainability}. \deleted{We also build on object-oriented metrics and use threshold-based approaches to analyze code quality and smells (see Section~\ref{sec:data}).}

\added{\textbf{Software Metrics Visualization} -- }Visualizations of software metrics have already been used to investigate source code quality---we cannot report all approaches here but give a brief overview. For instance, Emden and Moonen~\cite{van2002java} present a graph-based approach to automatically detect code smells and investigate how code smells can be integrated into a code inspection tool. Murphy-Hill and Black~\cite{murphy2010interactive} integrate a software metrics visualization with a source code view. 
Erdemir et al.~\cite{erdemir2011quality} provide a graph-based visualization tool encoding multivariate metrics in the node glyphs of the graph. Diverse software maps and cities~\cite{bohnet2011monitoring, Balzer2004Software, Wettel2011Software} 
arrange the structure of a software project on a 2D map and visually encode software metrics on top of this structure (often in the form of 3D buildings). 
Mumtaz et al.~\cite{mumtaz2018detecting} support the interactive detection of code smells using parallel coordinates plots and scatterplots.
%
Also, several tools are available that build on visualizations to assist developers in analyzing code quality in terms of software metrics. 
For instance, SonarQube~\cite{sonarqube}
controls and manages the code quality in several ways, such as continuous inspection and \deleted{tricky-}issue detection. The platform shows issues like code smells, bugs, etc. using\deleted{some} lightweight visualizations. It also helps developers\deleted{in} collaborating\deleted{the coding assignments} with a shared vision of code quality. There are some code analysis tools that accomplish similar tasks as SonarQube, for instance, Checkstyle~\cite{checkstyle}
and PMD~\cite{pmd}.
In contrast to these visualization approaches and tools, we focus on providing more context to the data\deleted{ presented} and explain the findings and their background in detail. We are not aware of any approach that uses a sophisticated text generation approach for this purpose.

\added{\textbf{Embedded Visualization} -- Visualizations included into the lines or paragraphs of a text are known as \textit{sparklines}~\cite{Tufte2006Beautiful}, \textit{word-sized}~\cite{beck2017word}, or \textit{word-scale} graphics~\cite{goffin2014exploring}. They allow close and coherent integration of the textual and visual representations of data. Some approaches apply these in the context of software engineering and embed them into the code to assist developers in understanding a program. Harward et al.~\cite{harward2010situ} and Sul{\'\i}r et al.~\cite{sulir2018visual} suggest augmenting the source code with visualizations to keep track of the state and properties of the code. Beck et al.~\cite{beck2013visual,beck2013situ} implement embedded visualizations for understanding program behavior and performance bottlenecks. Similarly, Hoffswell et al.~\cite{hoffswell2018augmenting} and Swift et al.~\cite{swift2013visual} augment source code with visualizations to aid understanding of runtime behavior. We embed visualizations into natural language text (not into source code) to support better understanding of the quality of the source code.}

\added{\textbf{Natural Language Generation} -- }Natural language generation allows us to automatically generate text from raw or pre-processed data~\cite{Reiter2000Building}. There are plenty of approaches that focus on automated text generation~\cite{Gatt2018Survey}. Few solutions also deal with the combined generation of text and visualization. Such combinations have been applied, for instance, to operational or maintenance-related instructions for mechanical devices ~\cite{Wahlster1993PlanBased}, explanatory captions of diagrams for information graphics~\cite{Mittal1995Generating}, reports of brain imaging data~\cite{Jordan2014TBIDoc}, weather forecast reports using predictive data analysis~\cite{Sripada2014Case, Ramos2015Linguistic}, evaluation of learning analytics in education~\cite{Ramos2017Evaluation}, and author profiles of the visualization community~\cite{Latif2018VIS}. Automated text generation has also been employed in the context of software engineering, for instance, for describing software models, such as class diagrams or use case diagrams, where they are augmented with textual descriptions~\cite{Lavoie1996ModelExplainer, Lavoie1997Customizable, Meziane2008Generating, Burden2011Natural}. Even more related to our solution are those approaches that deal with textual reports for source code, such as code documentation and summarization~\cite{Sridhara2010Towards, Moreno2013Automatic, McBurney2014Automatic}, commit messages~\cite{Cortes2014Automatically}, or reports on the runtime behavior of the code~\cite{Beck2017Method}. However, none of these approaches discusses code quality. Also, the results are usually not presented as highly interactive documents that support the exploration of the data. 

\added{\textbf{Text--vis Linking} -- }The interactive linking of text and visualizations has only been explored to some extent. Beck and Weiskopf~\cite{beck2017word} propose an abstract interaction model for documents containing text, word-sized graphics, and regular visualizations; all three types of data representations are linked via brushing-and-linking. 
Latif et al.~\cite{latif2018exploring} describe an authoring solution for web documents to produce some of those interactions. Our interaction model also uses and extends the model by Beck and Weiskopf. \added{Kim et al.~\cite{kim2018facilitating} advocate for \emph{text--text} linking to facilitate document reading. In their approach, the linking is supported between text in the main body and text in tables.} Few of the systems that generate both text and visualization---for instance, \textit{VIS Author Profiles}~\cite{Latif2018VIS} and \textit{interactive Map Reports}~\cite{latif2019imr}---discuss interactions, but still focus more on explanations and offer limited data exploration. \textit{Voder}~\cite{Srinivasan2018Augmenting}, in contrast, focuses more on interactions and supports the data exploration process by offering short descriptions about key findings in the data. However, it does not generate a comprehensive report with longer descriptions. 


In summary, although existing approaches present source code information, they lack in putting data into context and providing explanations. None of the systems, also outside the software engineering community, supports exploranation as a process blending explanations and explorations in a way that we envision leveraging the interactive combination of textual and visual descriptions. \added{We are inspired by the abstract idea of interactive linking of text and visualizations by Beck and Weiskopf~\cite{beck2017word} to support exploranation. We adopt CK, QMOOD, and McCabe's metrics (listed in Table~\ref{tab:metrics}) and use them in combination with pre-defined thresholds to analyze and present source code quality.} 

\section{Exploranative Documents for Multivariate Data} \label{sec:concepts}

With our approach, we want to support \emph{active reading}~\cite{adler1972read} of multivariate data. The readers should not be restricted to passively consume explanations but be facilitated to actively explore the data as well. They can read with a specific focus, break the content apart, and analyze its meaning. In contrast to traditional visual analytics interfaces, which focus on exploration, we provide more guidance to the users in the form of textual explanations, while---to some extent---preserving the explorative power of an interactive visualization. While our concrete implementation is tailored to software metrics and related code quality characteristics, our general approach applies to multivariate data in a broader sense and might even extend to wider classes of interactive documents. In this section, we introduce the generic concept, before the following sections present the details of the tailored solutions. 

We build on ideas from  Ynnerman et al.'s concept of \emph{exploranation}~\cite{ynnerman2018exploranation} and Victor's \emph{explorable explanations}~\cite{victor2011explorable}. Ynnerman et al. introduce \emph{exploranation}---a coinage stemming from the terms \emph{exploration} and \emph{explanation}---for scientific communication, for instance, to show visitors of a museum a visualization of an ancient mummy, guide them to interesting aspects but also let them explore it; related earlier work by Weiskopf et al.\cite{Weiskopf:2006:relativity} combines visual explanation and exploratory visual experiments for similar scenarios. 
We adopt this idea for a visual analytics scenario. Although our targeted professional audience is narrower, we also want to make the data easy to access. Some of Ynnerman et al.'s design principles also apply to our scenario, in particular, that \emph{explorative microenvironments} blend with \emph{signposted narratives}. With respect to the visual representations used, Victor's suggestions for \emph{explorable explanations} are even closer to our work as they also focus on interactive documents. We employ \emph{explorable examples} and \emph{contextual information}, as two of three suggested categories of explanations. \deleted{In both Ynnerman et al.' and Victor's work, however, the focus is on education: teaching the users about a domain and using data as examples to do so. In contrast, we flip these priorities and, first of all, want to support the users in analyzing the data, while providing the users options for educating themselves on the side.}

Our approach of exploranative documents for multivariate data comprises the following building blocks. 


\textbf{(I) Textual Explanations} -- The main feature that distinguishes our approach from most other visual analytics solutions is the use and integration of automatically generated textual explanations. We discern different types of explanations: (i)~\textbf{Data-driven explanations} summarize the data and results of data analysis (e.g., identification of patterns or clustering) while pointing to remarkable observations and giving examples. (ii)~\textbf{Educational explanations} provide background on the domain concepts reflected in the data. (iii)~\textbf{Methodological explanations} give details about how the analysis was performed and the reason why the system came to certain conclusions. While the data-driven explanations are the focus of the documents, the two other types provide important context. The users can obtain, on demand, for what parts of the data summary they require background information. 

\textbf{(II) Explorable Visualizations} -- In addition to the textual explanations, the explorative component is mostly contributed by interactive visualizations. (i)~\textbf{Overview visualizations} should have a consistent location in the interface and be visible all the time. In the specific solution, we will build on parallel coordinate plots and scatterplots as such overview visualizations, but the approach is open to any appropriate visualization of multivariate data (e.g., scatterplot matrices, tabular-like representations, multivariate glyphs, projection methods). (ii)~\textbf{Embedded detail visualizations}, in contrast, enrich the text with further information and just show subsets or aspects of the data. There can be regular visualizations that scroll with the respective text, but also word-sized representations embedded in the text (\emph{sparklines}~\cite{Tufte2006Beautiful, beck2017word}). The better these visualization are integrated with the text, the easier it will be for users to explore them along reading the text. In general, the exploration process can happen visually by users deciding to look at and investigate certain elements of the visualization. On top of that, interactions to subselect the data and pick out individual elements further extend the users' abilities with respect to exploration.

\textbf{(III) Consistent Linking} -- Like in a traditional multi-view visual analytics system, a challenge is to maintain a clear linkage between the different views. However, in our case, the linking becomes even more difficult because data is described on very different levels of abstraction and with different modalities (text and visualization). We apply the concept of (i)~\textbf{vis--text interaction}~\cite{beck2017word}, which suggests linking all three representations---textual, visual, and embedded visual ones---in a bidirectional way using hover and click interactions. For instance, hovering a word in a text, the related entities can be highlighted across all views and also in textual descriptions. (ii)~\textbf{Consistent color coding} is used to further clarify relationships between the different textual and visual descriptions of related data. We suggest applying a consistent color coding of the different variables across all representations. Similar variables can be grouped by hue and \added{get} assigned a different brightness.

With these concepts, we support active reading and the exploranation process. The users are first confronted with a summary text and associated overview visualizations. One group of readers, especially first-time users, might follow mainly the provided narrative and start reading at the top left. Whenever something is unclear from the high-level summary, they can explore the required background on demand. After having read the main text, they might switch to exploring the data further using visualizations. In contrast, another group of users that might be more experienced could immediately start the data exploration process. While some information can be directly gained from the visualizations, for other insights occasionally reading the textual explanations can provide support. The textual summaries might also point them to interesting findings that they might have missed otherwise.

The resulting approach can be classified as a visual analytics solution that puts emphasis on presentation, storytelling, and dissemination. In terms of the sensemaking process described by Pirolli and Card~\cite{Pirolli2005Sensemaking}, it covers the sensemaking loop (i.e., build a case, tell a story, search for support, and reevaluate) rather than the foraging loop (i.e., search and filter, read and extract, search for information, and search for relations). 

\begin{table*}[tbp]
\centering
\caption{Class-level software metrics (name and acronym) used for code quality analysis, grouped by quality attributes.}
\label{tab:metrics}
\ssmall
\sffamily
\begin{tabularx}{\textwidth}{p{2cm}lp{2cm}L}
\toprule 
\textbf{Quality Attribute} & Software Metric & Acronym & Description \\
\toprule 
\textbf{Complexity} & Weighted methods per class & wmc & The sum of all method complexity values for a class. \\
& Maximum cyclomatic complexity & max\_cc & The maximum of all the method-level complexity values of a class. \\ \midrule
\textbf{Coupling} & Afferent coupling & ca &  The number of other classes that depend on a class (incoming dependencies). \\
& Efferent coupling & ce &  The number of other classes on which a class depends (outgoing dependencies). \\ \midrule
\textbf{Cohesion} & Lack of cohesion of methods & lcom3 & It checks whether the methods access the same set of variables of a class. \\ \midrule
\textbf{Inheritance} & Depth of inheritance & dit & \deleted{It measures} The inheritance levels for a class. \\
& Number of children & noc &  The number of immediate descendants of a class. \\ \midrule
\textbf{Other} & Average method complexity & amc & \deleted{It measures} The average size of the methods in a class. \\
& Lines of code & loc &  \deleted{It represents} The total lines of code present in a class.\\ & Number of public methods & npm &  The number of methods declared as public in a class. \\ 
& Number of bugs & bug &  \deleted{It counts}The \added{number of} bugs that have been associated with a class. \\ 
\bottomrule
\end{tabularx}
\end{table*}

\section{Code Quality Analysis} \label{sec:infoneeds}

In this section, we discuss the application scenario and explain the data processing required for it. 

\subsection{Targeted Users} \label{sec:targeted_users}

The targeted users of our system are mainly the stakeholders who are concerned with source code and its quality. Being well-informed about the quality of source code can help them in taking steps to improve various aspects that need attention. 
The group includes mainly software developers who work with the code on a daily basis; they need to read and understand the code to be able to extend and maintain it. But the group also extends to testers, software architects, product owners, or project managers. For example, stakeholders like product owners and project managers can use our system to assess the overall quality of a project and use the gained information to prioritize quality concerns. In essence, this target user group is much wider than just code quality experts and highly experienced developers. For this reason, we believe that integrating guidance and explanations will provide valuable support for most of the users and allows them to draw actionable conclusions. However, just presenting a static report would not suffice because the prepared summaries and explanations can only be a starting point for investigating a detected problem in detail. 

\subsection{Data} \label{sec:data}

Software metrics provide significant information on code quality\deleted{ because, usually, the measurement of quality attributes is accomplished using them}. We employ software metrics belonging to object-oriented metrics known as Chidamber and Kemerer metrics suite~\cite{chidamber1994metrics}\added{,} \deleted{and the} QMOOD metrics~\cite{bansiya2002hierarchical}\added{, and McCabe's complexity metric~\cite{mccabe1976complexity}}. The metrics are at class-level \added{abstraction}\deleted{, which is a mid-level of abstraction}; they quantify \added{quality} attributes of the classes of a software project. Specifically, we work with a subset of 11 metrics in total (Table~\ref{tab:metrics}). \added{The selected metrics are often employed to measure quality attributes, for instance, the \emph{coupling between objects} metric measures the degree of interdependence between classes, which is consistent with how coupling is traditionally defined. Excessive coupling results in a weak modular design and limits the maintainability of a class. This means that the coupling properties of a class need to be analyzed to ensure better modularity and maintainability. Similarly, other metrics measure different characteristics of the classes that need to be monitored for better software quality.} We selected the \added{object-oriented} metrics \replaced{that}{in order to} quantify \added{four} different quality aspects\replaced{:}{,} \deleted{such as}complexity, coupling, cohesion, and inheritance. \added{The metrics listed in ``Other'' category (in Table~\ref{tab:metrics})} reflect general properties like size or are required for the detection of code smells. To show how bug-prone the code has been, we include the \emph{number of bugs} associated with each class at a respective point during the development process.

\subsection{Code Quality Analysis} \label{sec:analysisGoals}



\deleted{We identify quality aspects that correlate with source code by means of scenarios and data processing.}From the related work, we observe that there are many \deleted{aspects}\added{class-level metrics} that can be linked to code quality~\cite{ouni2013maintainability, mumtaz2018detecting, stamelos2002code}. These approaches employ\deleted{software metrics with} thresholds to \replaced{detect}{measure} code quality \replaced{issues}{aspects}. Software metrics are also used to express abstraction level quality characteristics, such as coupling, complexity, cohesion, and inheritance~\cite{filo2015catalogue, shatnawi2010finding}. In our system, we also apply software metrics with the thresholds defined by Fil\'{o} et al.~\cite{filo2015catalogue} to measure these quality attributes. \deleted{While analyzing the software metrics, we recognize various ways}\added{We rate} the quality level (good, regular, or bad) of a component or the severity level of a problem (high, medium, or low). \deleted{in the project is expressed}\deleted{We use such classification to segregate the classes in term of these quality levels. The first category tags classes with good quality; the second category lists classes with regular quality; and the third category has classes with bad quality.}

\deleted{Code smells can also be connected to software metrics.} Code smells provide information on implementation decisions or choices that might degrade code quality~\cite{zhang2011code}. Again \replaced{based on}{from the support of the} related work~\cite{ouni2013maintainability},\deleted{we identify parts of the source code that can lead to technical debt because of the presence of code smells.} we detect four types of common class-level code smells: \emph{Large Class}, \emph{Functional Decomposition}, \emph{Spaghetti Code}, and \emph{Lazy Class} using class-level metrics. \emph{Large Class} is the one that \replaced{has many fields and methods, resulting in many lines of code}{handles most of the system's functionality}~\cite{zhang2011code}. A class with many private fields and methods is associated with \emph{Functional Decomposition}~\cite{zhang2011code}. A class with \emph{Spaghetti Code} has long methods \replaced{without proper structure}{with many conditional statements}~\cite{zhang2011code}. A class with little to no functionality is a \emph{Lazy Class}~\cite{zhang2011code}. Since we have class-level metrics\deleted{data}, it is possible to compute these code smells using predefined thresholds.\deleted{ We analyze the detection rules of the code smells to investigate the working and correlation of software metrics. The detection rules share the software metrics with different threshold values to classify a certain code smell. We base our educational and methodological explanations on such analysis of software metrics and their thresholds. Furthermore, we use software data analysis in the context of code quality to generate relevant data-driven explanations.}

\deleted{We also discuss bugs in the software projects because they also correlate with code quality. We focus our analysis to those software classes that have code smells and also prone to bugs. The goal of the analysis is to spotlight those classes that are potentially more quality deterrent. We express these details in the data-driven explanations presented in the quality documents.}

Based on \added{the metrics and} this analysis, the content of the code quality document comprises three parts: first, quality attributes covering coupling, complexity, cohesion, and inheritance; second, code smells in terms of \emph{Large Classes}, \emph{Functional Decomposition}, \emph{Spaghetti Code}, and \emph{Lazy Classes}; and third, information about bug history.

\begin{figure}[tb]
 \centering 
 \includegraphics[width=1.0\columnwidth]{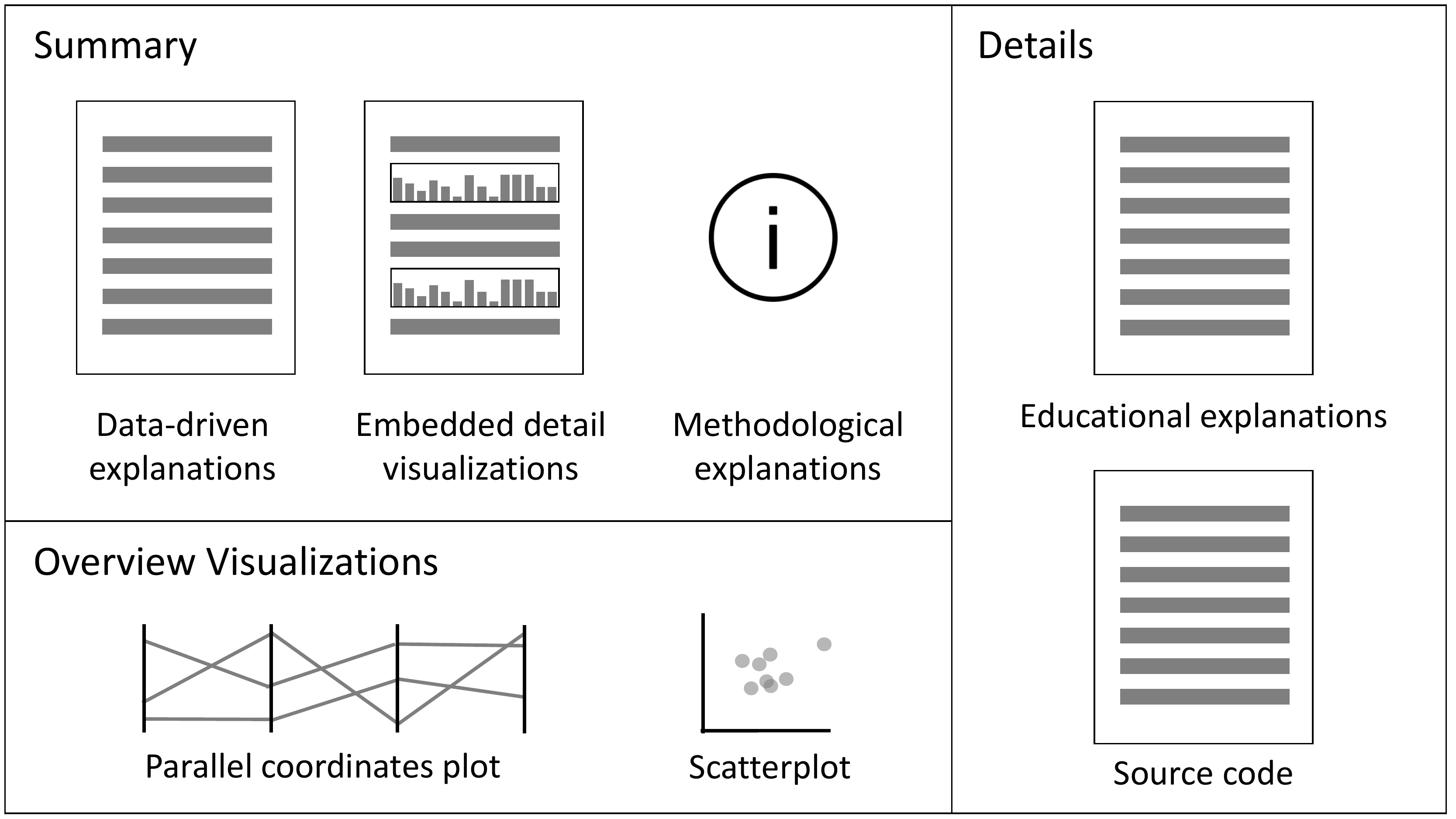}
 \caption{Abstract representation of the interface structure.}
 \label{fig:interface}
\end{figure}

\begin{figure*}[tb]
 \centering
 \includegraphics[width=1.0\linewidth, trim=0.0in 0.0in 0.0in 0in]{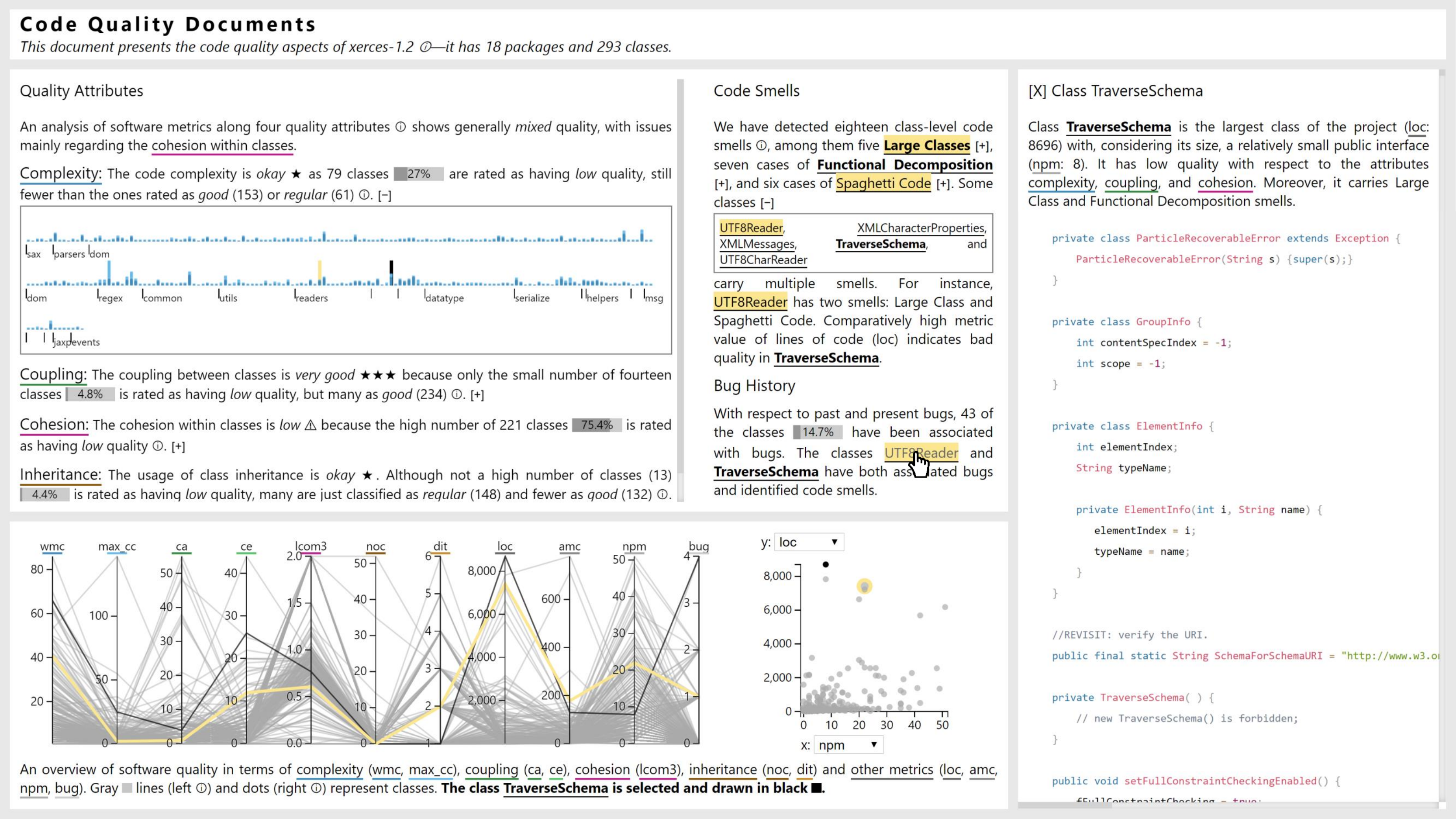}
 \caption{Various instances of \emph{vis--text} interactions. A persistent highlighting (click on \emph{TraverseSchema}) marks the related elements with bold font in text (\emph{text--text}), a black line in the parallel coordinates, a black dot in the scatterplot (\emph{text--vis}), and black background in embedded detail visualizations (\emph{text--emvis}). Similarly, a non-persistent highlighting (hover on \emph{UTF8Reader}) marks the corresponding elements in yellow. The details panel shows the class-specific description and source code of the persistently selected class, \emph{TraverseSchema}.}
 \label{fig:interaction1}
\end{figure*}

\section{Exploranative Code Quality Documents} \label{sec:approach}

\deleted{Based on}\added{Using} the described software quality analysis (Section~\ref{sec:infoneeds}) and our generic approach (Section~\ref{sec:concepts}), we have developed a visual analytics solution named \emph{Code Quality Documents}. Our system is implemented as a web application using the libraries \emph{D3js}, \emph{JQuery}, \emph{JQuery Sparklines} and builds on the standard \emph{D3js} implementations of parallel coordinates plots and scatterplots. In this section, we first give an overview of the interface, then explain the individual components, and finally describe the interaction model that links the different components.

\subsection{Interface Structure}

We designed a multi-view interface with different panels for overview descriptions and details. Figure~\ref{fig:interface} shows an abstract representation of the interface, which maps the building blocks described in Section~\ref{sec:concepts} to the different panels---specific examples of the interface can be found in Figure~\ref{fig:teaser} and Figure~\ref{fig:interaction1}. The summary panel presents the main data-driven textual explanations~(I.i) summarizing the results of code quality analysis (Section~\ref{sec:analysisGoals}). In addition, it contains embedded visualizations~(II.i) and methodological explanations can be retrieved on demand~(I.iii). The panel for the overview visualizations~(II.i) contains a parallel coordinates plot and a scatterplot. The details view provides educational explanations~(I.ii) or class details on selection. 


\begin{figure}[tb]
 \centering 
 \fbox{\includegraphics[width=0.98\columnwidth, trim=0in 3.7in 6.8in 0.1in]{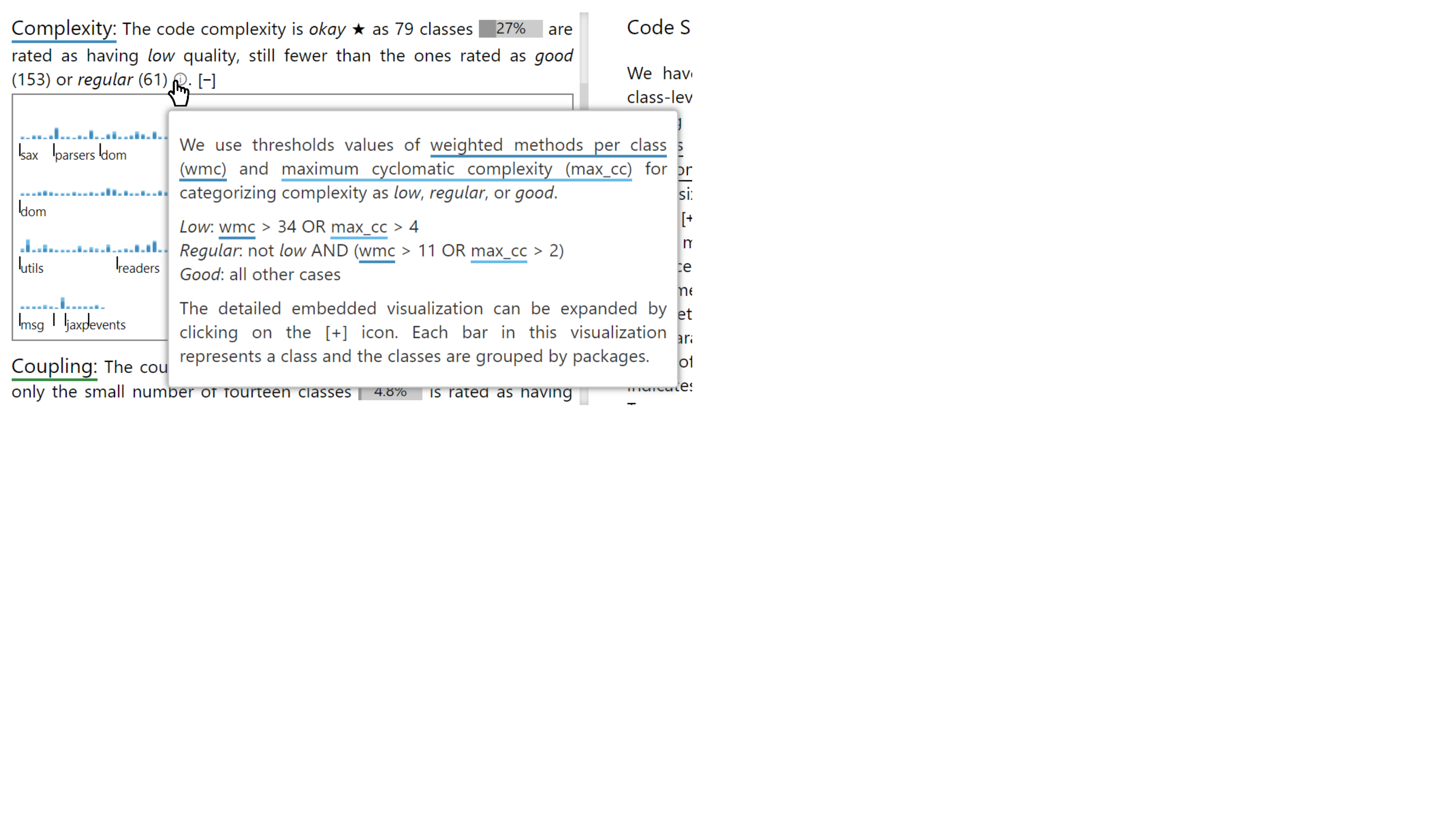}}
 \caption{Methodological explanation of classifying complexity in terms of low, regular, or good.}
 \label{fig:interaction2}
\end{figure} 

The data-driven text describes code quality along with quality attributes, code smells, and bugs. The overview text contains detailed embedded visualizations and a list of classes corresponding to certain categories of code smells. By default, these details are not expanded in this panel. An icon [+] indicates that details are available and can be expanded. \added{This dynamic expansion is similar to the concept of interline in fluid documents~\cite{zellweger2000impact}, where additional details are shown within the lines of the text.} An info icon {\large\textcircled{\small i}} hints at the presence of methodological explanations. Hovering this icon presents a tooltip \added{(similar to a popup in fluid documents~\cite{zellweger2000impact})} that describes the methodology used to come up with the respective detail. In Figure~\ref{fig:interaction2}, the tooltip shows the software metrics and their thresholds used to classify low, regular, or good complexity. \added{We argue that highlighting different quality issues with methodological explanations can assist users in making decisions to improve the quality.}

Educational explanations are related to the domain of code quality. The names of the quality attributes and the software metrics carry a thick colored border at the bottom. 
Clicking on these terms brings up background information in the details view. The description also includes project-specific examples to better communicate the concept. For instance, Figure~\ref{fig:teaser}{\large\textcircled{\small D}} shows the educational explanation for the quality attribute \emph{cohesion}; \emph{WhitespaceAnalyzer} is provided as an example of a class having the highest value of \emph{lack of cohesion of methods}. In a similar way, explanations on code smells can be accessed. Clicking on a class anywhere in the system opens the source code of that class in the details view preceded by a class-specific explanation providing a short summary of problems in the class, if any.

\subsection{Natural Language Generation}

To automatically produce textual descriptions, we need to generate text from data. We aim at a self-explanatory and understandable text that has a natural flow and appears almost as if written by a human. To this end, we employ template-based natural language generation~\cite{Gatt2018Survey}, a simple, but effective and easy-to-use generation method. 

\added{Our text generation process is based on previous work~\cite[Section IIIA]{Beck2017Method}\cite[Section 4.2]{Latif2018VIS} and can be modeled as directed decision graphs---one for each subsection of the report. The decision graph consists of \textit{start}, \textit{stop}, \textit{decision}, and \textit{text} nodes. Any traversal of the graph from \textit{start} to \textit{stop} node results in a report section. \textit{Decision} nodes control the flow of the graph based on the values of decision variables.} \replaced{\textit{Text} nodes}{Templates} consist of predefined templates with variables in them filled with the information coming from data. \added{Traversing a \textit{text} node adds a phrase or sentence to the report section. Even in the \textit{text} nodes,} we make extensive use of conditions, not just to account for different grammatical cases such as plural or singular but also to describe different cases and provide reasons for an analysis result. For instance, we not only list the number of classes that have low quality with respect to any of the four quality attributes but explain reasons for the rating. The same rating can even have different reasons (Figure~\ref{fig:interaction1}: \textit{``The code complexity is \emph{okay} as 79 classes are rated as having low quality, still fewer than the ones rated as \emph{good} (153) or \emph{regular} (61).''} and \textit{``The usage of class inheritance is \emph{okay}. Although not a high number of classes (13) is rated as having low quality, many are just classified as regular (148) and fewer as good (132).''} We leave out sentences if no results are available and handle special cases.
For instance, a special case for this paragraph is when no code smells were found: \textit{``We have not detected any class-level code smells in the project---congratulations.''} 

\subsection{Visualizations} \label{sec:visualizations}
For data exploration and context of the textual descriptions, we provide two overview visualizations: a parallel coordinates plot and a scatterplot. These visualizations are useful in discerning important patterns and relationships between metrics~\cite{schulz2017visual}. We argue that users can find these visualizations \added{useful}\deleted{helpful} in\deleted{better} understanding the code quality and comparing the properties of different classes. To obtain an overview of all the metrics, the parallel coordinates plot is helpful, whereas the scatterplot supports the identification of relationships between two metrics.

In addition to these visualizations, the embedded detail visualizations complement the text generated in the document. We employ small bar charts to represent the metric values of one category for all classes, structuring the classes with respect to the packages in which they are contained. Furthermore, we use word-sized bar charts to indicate percentage values. The values always refer to problematic cases (e.g., low quality or bug-prone classes) and are given relative to the overall number of classes. Small star icons and warning symbols provide a quick hint of the respective rating for each quality attribute. 

We use consistent color coding to couple the visualized metrics. For instance, in Figure~\ref{fig:interaction1}, the colors of complexity metrics in the caption of the parallel coordinates plot match the color coding of the complexity bar chart in the quality attributes section. 
The metrics are grouped in terms of the quality attributes---two metrics of the same group are associated with the same hue but a different brightness. 

\subsection{Interaction Model}

\begin{figure}[tb]
 \centering 
 \includegraphics[width=0.98\columnwidth, trim=1.6in 0.5in 0.9in 0.4in, clip=true]{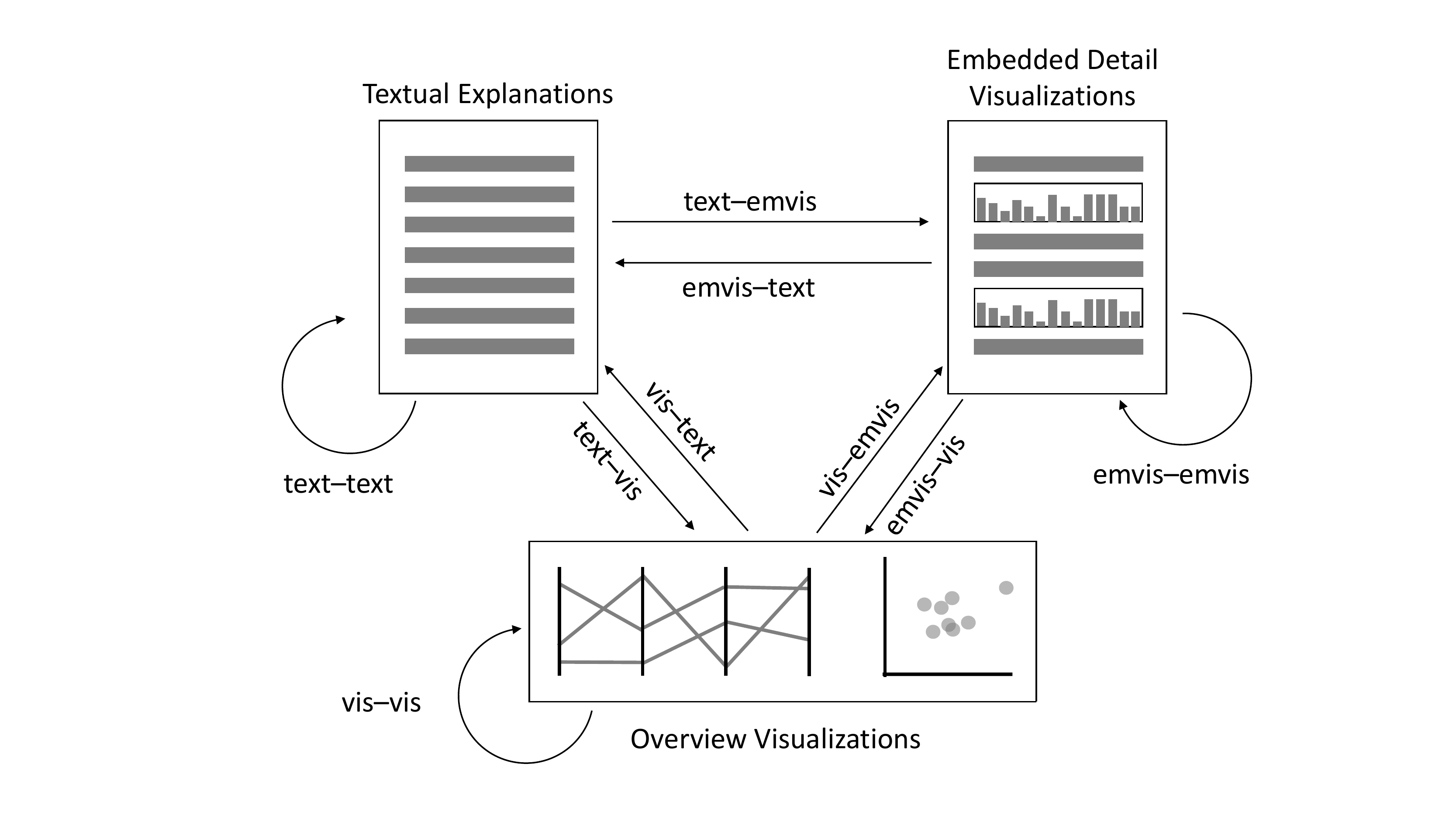}
 \caption{Abstract representation of the interaction model. We have bi-directional interactions among text, embedded visualizations (emvis), and overview visualizations (vis).}
 \label{fig:interaction}
\end{figure} 

\begin{figure*}[tb]
 \centering 
 \fbox{\includegraphics[width=0.98\textwidth, trim=0in 5.1in 4.9in 0in, clip=true]{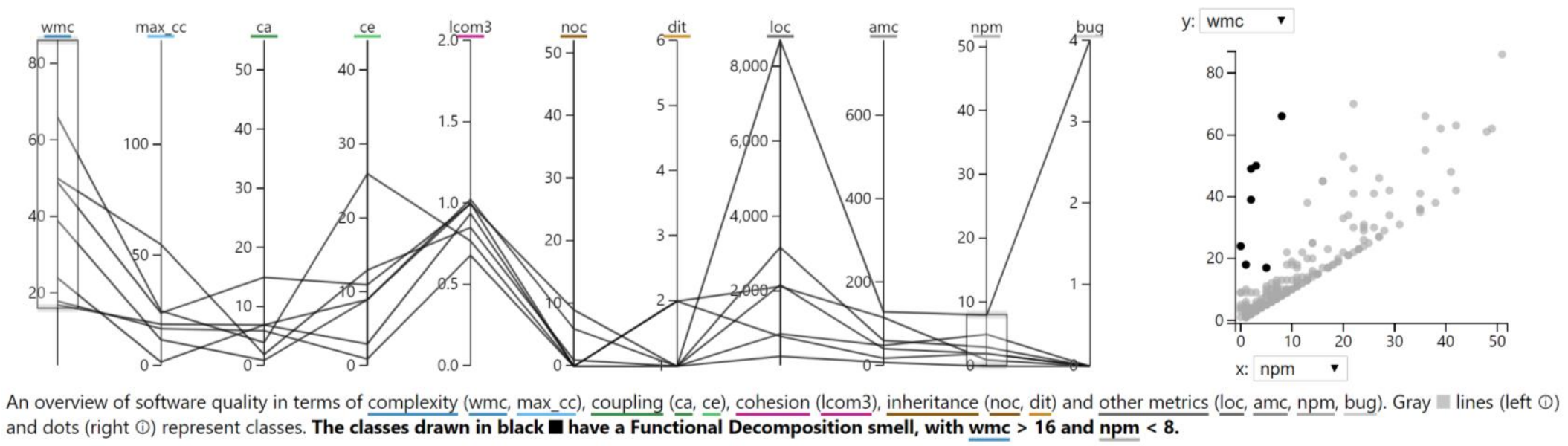}}
 \caption{A selection of the classes in \added{\emph{Xerces 1.2}} that have a \emph{Functional Decomposition} smell; they are highlighted in the parallel coordinates plot and the scatterplot. The caption of the visualizations adapts to describe the selection.}
 \label{fig:captions}
\end{figure*} 

To connect textual and visual descriptions\deleted{ with each other}, consistent interactive linking is important (see also Section~\ref{sec:concepts}). We build on the concept of \emph{vis--text} interaction introduced by Beck and Weiskopf~\cite{beck2017word}. In addition to the local interactions discussed above (i.e., interactions that only locally impact the interface, such as details blended in on demand), a global interaction model is intended to link the various visual representations. In slight adaption to the model of Beck and Weiskopf, we discern the textual explanations (\emph{text}), overview visualizations (\emph{vis}), and embedded visualizations (\emph{emvis}). As shown in Figure~\ref{fig:interaction}, these representations are interactively linked to each other in a bi-directional way. 

As these components contain different representations of the same class-level multivariate data, an essential interaction is the brushing and linking of data points across all representations. Hovering over a class name anywhere in the text triggers a transient selection (a non-persistent selection). It highlights the corresponding poly-line in the parallel coordinates plot (\emph{text--vis}), the dot in the scatterplot (\emph{text--vis}), the bars in the embedded visualizations (\emph{text--emvis}), and other occurrences of the class name in the text (\emph{text--text}). Figure~\ref{fig:interaction1} shows the effect of hovering over the class name~\inlinegraphic{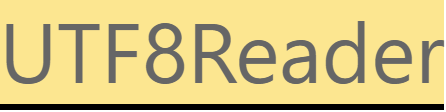}; the linked parts are highlighted with yellow color. 
Apart from the other instances of the hovered class in text, we also mark the corresponding code smells. \added{The \emph{Large Class} and \emph{Spaghetti Code} are highlighted since \emph{UTF8Reader} contains both of these code smells (Figure~\ref{fig:interaction1}).}
Hovering over a bar in the embedded visualization and a dot in the scatterplot has a similar effect and triggers \emph{emvis--vis}/\emph{vis--vis}, \emph{emvis--text}/\emph{vis--text}, and \emph{emvis--emvis}/\emph{vis--emvis} interactions.

The transient selection shows as long as the interactive element is hovered and provides a quick way of cross-referencing different representations. To make the highlighting persistent, the interactive elements can be clicked; the parts related to the clicked element are highlighted with black color in the visualizations and with bold font in the text. 
For instance, Figure~\ref{fig:interaction1} shows the persistent selection corresponding to the class~\inlinegraphic{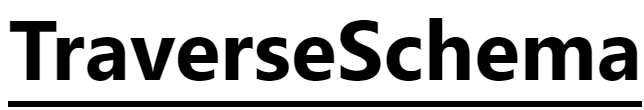}. This helps in getting a comparative overview of two different classes with respect to various aspects\replaced{;}{. For instance, } one glance at Figure~\ref{fig:interaction1} is sufficient to tell that \textit{UTF8Reader} and \textit{TraverseSchema} have one code smell (\emph{Large Class}) in common. In addition, we can quickly observe that \emph{UTF8Reader} has less complexity (embedded visualization for complexity), fewer \emph{lines of code} (scatterplot), and fewer bugs (parallel coordinates plot) than \emph{TraverseSchema}.

Clicking on a code smell\deleted{ name}, aside from showing an educational explanation in the details view, highlights the set of classes that contain that code smell in the parallel coordinates plot and scatterplot (\emph{text--vis})\added{---Figure~\ref{fig:captions} shows the result of clicking \textit{Functional Decomposition}}. This helps in understanding the pattern of metric values for the classes having different code smells. Since the scatterplot illustrates the relationship between any two software metrics, we update the dimensions of the scatterplot on persistent interactions according to the context. For example, clicking on \emph{Functional Decomposition} will update the scatterplot dimensions to \emph{weighted methods per class} and \emph{number of public methods} as these metrics are used to identify the smell (see Figure~\ref{fig:captions}). Moreover, users can explore the relationships between \replaced{other}{any other pair of} metrics. Since lines are hard to select in a parallel coordinates plots (they are thin and often occlude each other), we provide a persistent range selection on the axes (brushing interaction with mouse press and hold). On every persistent selection, the caption of the figures adapts accordingly to describe the selected elements (see Figure~\ref{fig:captions}). \added{In contrast to legends, the textual captions allow for the inclusion of contextual and methodological information (e.g., data filtering criterion), which helps in making the interactive visualizations more self-explanatory.}


\section{Design Process and Evaluation} \label{sec:evaluation}


We followed a design process in which the authors of this paper worked in close collaboration as a team. Initially, the team was composed of three members, but later another member was included because of a specific skill-set. The fourth member was included in the fourth quarter of the design process. The team has a blend of different backgrounds and expertise. One member has expertise in software engineering; one member works in the domain of software visualization; the third has experience in visual analytics and visualization in general; the fourth member (included later) has skills in automated text generation and interactive documents. The team members were located at different institutes, therefore, frequent visits were arranged. One member spent a week at a lab where other members usually collaborate. In addition, we organized regular meetings to share ideas and discuss the outcomes of the design decisions for our system. Some of the implementation work was completed in a pair programming setting. The process went through a period of approximately 9 months with a focus on designing a visual analytics system for a domain-specific problem.

Along the design process, we conducted a formative evaluation (i.e., an evaluation that focuses on testing and improving intermediate prototypes)---in the fourth quarter of the design process---to validate the problem analysis and our visual analytics solution. We invited participants who were not part of the design process team \added{and are not authors of this paper}. From our formative study, we obtained user feedback and observed usage strategies. \added{We performed a qualitative analysis of the participants' feedback---received through a questionnaire---where they expressed their views (positive and negative) about our system. In the questionnaire, we formulated tasks to analyze the usability and usefulness of our system. In particular, through our formative study, we assessed how different views coordinate with each other; how \emph{vis--text} interaction support the exploration process; how text, visualizations (embedded and non-embedded), and their interactive linking help in the code quality analysis process.} The results helped us refine and improve our system. We conducted two iterations within the formative study. The data from the user study is provided in the supplemental material\cite{dataAndTool}.



\textbf{First Iteration} -- In terms of textual explanations, the first tested version of the system had data-driven explanations, some educational explanations, but no methodological explanations. From the interaction perspective, the system had a few one-way \emph{text--vis} interactions and a few \emph{emvis--emvis} interactions. Consistent linking was implemented minimally. The first iteration of the formative study was organized with four participants (3 PhD students and a postdoc), two with a software engineering background and the other two working in visualization. We decided for this mix of experts to receive educated feedback both on the content of the documents as well as the data presentation. The study included three phases and took approximately 45 minutes per participant. In the first phase, participants were asked to use the system, identify different aspects of code quality discussed in the document, and summarize them. The second phase involved reviewing different features of the system and providing detailed feedback on each.
As the third phase, we interviewed the participants, asked general questions, and concluded the session. \added{We performed a qualitative analysis of the participants' feedback. Although we received quantitative feedback (through Likert scale) as well, we were more interested in textual responses because they helped us identify areas for improvement. 
We summarize the results as follows, also discussing our improvements that we implemented as a reaction to the feedback.}
 
All participants found the data-driven explanations helpful and complementing the visualizations \added{(\emph{``It is nice to see discussion on some quality aspects of code. The tool analyzes the data well; interactions and details of code are useful.''})}. However, due to the lack of educational and methodological\deleted{related} explanations, they were unclear about certain descriptions \added{(\emph{``I have [had] a hard time trying to figure out the meanings of the acronyms [metrics] in parallel coordinates plot and scatter plot [...]''})}; two participants disagreed that the document is self-explanatory \added{due to lack of}\deleted{ because we did not provide} educational and methodological explanations (\emph{reaction}: we consistently added educational and methodological explanations). One participant argued to exclude details about the project from the main text because it distracted the focus from the aspects discussed in the document (\emph{reaction}: we moved this text to the header to clarify that it is meta information and not a regular part of the document).  

The participants suggested to have more interactions to make the system more intuitive\deleted{to use}. The\deleted{main} missing interactions were \emph{vis--text} and \emph{emvis--text} \added{(\emph{``I did not find a link from the visualization to the text. But the other way round [it] works nicely.''})}. Although we had some \emph{text--vis} interactions present in the system, it was interesting that participants were expecting\deleted{to be able} to click on visual representations of classes. The scarcity of interactions was hindering the exploration process. They also pointed out the lack of \added{\emph{emvis--vis}} linking\deleted{between embedded and overview visualizations} (\emph{reaction}: we systematically added interactive linkage between all representations of the same data). 
The participants emphasized that the overview visualization should not scroll out\deleted{scroll-bar} but be visible at all times \added{(\emph{``The placing (initially 60\% off-screen, scrolling is needed) is bad [...]''}}; \emph{reaction}: we moved the overview visualizations to a separate panel).

We received mixed responses on the parallel coordinates plot and scatterplot, but the participants did not use the visualizations for exploration yet \added{(\emph{``Maybe they [parallel coordinates plot and scatterplot] would be useful to assess some hypotheses, but for exploratory analysis it is not clear how to use them [...]''})}. We assume the main reason was a lack of interactive linking, especially between the two visualizations and with the textual explanations (\emph{reaction}: we consistently added such interactive links). The participants also suggested to include legends and captions in the visualizations to make them self-explanatory because sometimes they were unable to understand the changes happening in the visualizations (\emph{reaction}: we added captions and\deleted{carefully} provide legends as part of the captions and in the details panel).


\textbf{Second Iteration} -- In the second iteration of our formative evaluation, the system incorporated all types of textual explanations, however, still missing a few explanations. Furthermore, most of the transient interactions through consistent linking of text and visualizations were provided, without the full implementation of the persistent interactions yet. We invited three participants (two PhD students and one postdoc) who participated in the first phase to identify improvements in the system and, for a fresh perspective, one new participant (PhD student) to evaluate the system. The new participant has a background in software engineering and is currently conducting information visualization research. We followed a similar procedure as in the first iteration, however, removing the first phase because we did not expect new insights if repeating it; the duration of the study was reduced to about 30 minutes per participant. We also updated the interview questions to focus on the specific improvements and the general applicability of our approach. 

\begin{figure*}[tb]
 \centering 
 \fbox{\includegraphics[width=0.98\textwidth, trim=0.05in 3.3in 0.45in 0.05in, clip=true]{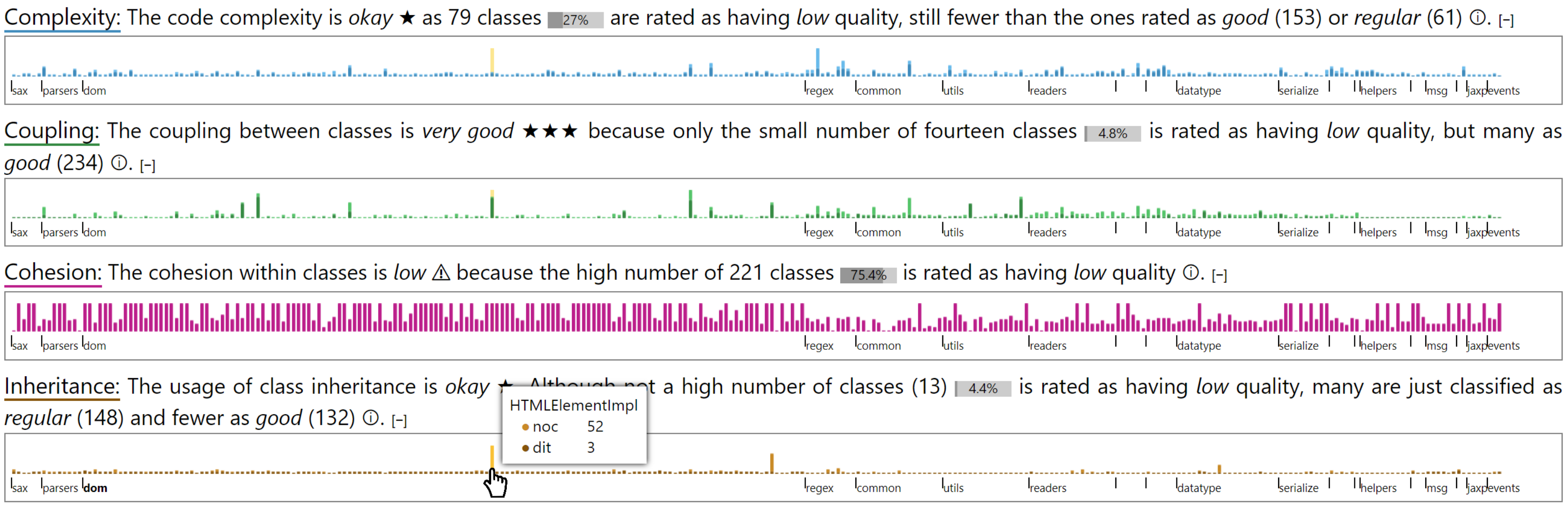}}
 \caption{Low quality in \emph{HTMLElementImpl} found through active exploration of the embedded detail visualizations of coupling, cohesion, and inheritance. 
 }
 \label{fig:interaction3}
\end{figure*} 

All participants agreed that the system was improved overall. We received positive feedback on the layout, textual descriptions, coordinated views, interactions, and linking \added{(\emph{``The layout is more compact. I can see all the information without having to scroll, which benefits the analysis when highlighting elements of the coordinated views. The interface looks clean, which facilitates to concentrate on the visualizations.''})}. The participants confirmed that the system supports exploration well and textual explanations are useful and well connected. However, one of the participants suggested to include more information about bugs (\emph{reaction}: \emph{none}---we do not have access to more details; retrieving these and extending the description accordingly remains future work). 
All participants, except one, expressed that the document is self-explanatory. He wanted more methodological and educational explanations on metrics \added{(\emph{``[...] more information why certain metrics are grouped together and what are the background of those [metrics].''}}; \emph{reaction}: we added more detailed explanations on metrics). 
It was also unclear to him that labels under the small bar charts are package names (\emph{reaction}: we now provide an explanation of the bar chart in the respective methodological explanation, see Figure~\ref{fig:interaction2}). 
%
One participant did not find horizontal embedded bars (that show percentages) useful because they do not add much to the text (\emph{reaction}: we reduced the number of shown bars and improved their consistency; they now always refer to problematic classes as a fraction of the total number of classes). The majority of the participants accentuated that the system can\deleted{certainly} be useful for educational purposes (e.g., teaching students about \added{code-related issues}\deleted{the issues in the code}). 

\added{Despite incrementally improving, some limitations remain in our visual analytics system. Currently, the system analyzes the software quality but does not explicitly guide the users in correcting measures. For instance, it discusses different code smells in a project without suggesting suitable refactorings. We also could not include more information about bugs besides \emph{number of bugs} because finding access to relevant information (e.g., bug appearance timeline, active duration, resolution) is difficult. Although we have optimized for self-explanation, misinterpretation of textual descriptions and visualizations is still possible. However, through the interactive linking of text and visualizations, we hope potential misunderstanding can be mitigated. In terms of evaluation, the participants had appropriate knowledge of visualization and code quality to provide substantial feedback on visualization design, usability, and code quality analysis. Additional valuable insights could be obtained through a larger-scale user study with an appropriate blend of targeted users having industrial experience.}

\section{Application Examples} \label{sec:examples}

With two application examples, we demonstrate how our exploranative documents \added{can} assist in analyzing code quality. This section also implicitly reflects on how users can work with our interactive document and the support they receive from it to achieve various analysis tasks. We demonstrate the working of our exploranative code quality documents using two software projects: \emph{Xerces~1.2} (an XML handling framework)~\cite{lucene,lucene2010Data}
and \emph{Lucene~2.0} (a search library)
\cite{xerces,xerces2010Data}. We demonstrate how textual explanations point to a specific issue in the code, and how exploration helps in analyzing problems. 

In the \emph{Xerces 1.2} project, our system identifies 18 code smells (Figure~\ref{fig:interaction1}). From the textual explanation, we read that \emph{UTF8Reader} has \emph{Large Class} and \emph{Spaghetti Code}. \emph{UTF8Reader} becomes more interesting when we realize that this class has had a large number of bugs as well. Looking into the code of the class, we find a confirmation of these issues self-admitted by the developers in the class description comment, also providing a reason: ``\emph{Some blatant v[io]lation of good object oriented programming rules, ignoring boundaries of modularity, etc., in the name of good performance.}'' \added{This is an example that shows that our approach helps the users focus on most problematic classes.}

Another class highlighted in Figure~\ref{fig:interaction1} is \emph{TraverseSchema} because it also carries two code smells (\emph{Large Class} and \emph{Functional Decomposition}) and has a history of bugs. In the class description, we read that \emph{TraverseSchema} is the largest class in the system and its quality is low with respect to complexity, coupling, and cohesion (see Figure~\ref{fig:interaction1}, details panel). Exploration through consistent linking helps locate this class in the parallel coordinates plot and the scatterplot (see Figure~\ref{fig:interaction1}, black line and dot); it shows a similar pattern as \emph{UTF8Reader} (see Figure~\ref{fig:interaction1}, yellow line and dot). Investigating the source code of \emph{TraverseSchema}, we confirm the existence of a \emph{Functional Decomposition} smell because it has several private but few public methods. 

During an exploration of the bar charts in the quality attributes section, we find that \emph{HTMLElementImpl} has a complex inheritance structure indicated by an extremely high \emph{number of children} (Figure~\ref{fig:interaction3}; 52 children, while the threshold of low quality is only more than three children). The extending classes are also part of the \emph{dom} package and can be investigated through the bar chart. We head on to explore the quality of \emph{HTMLElementImpl} from other quality attributes as well. We notice that the related metrics of coupling and cohesion are also high, meaning low quality in these attributes, however, the complexity metrics are in acceptable range (see Figure~\ref{fig:interaction3}). Similar information is also mentioned in the summary presented with the source code. 

\added{We use the parallel coordinates plot and the scatterplot to learn more about outliers and the interplay of metrics. For instance, in Figure~\ref{fig:captions}, a positive correlation can be observed between \emph{weighted method per class} (wmc) and \emph{number of public methods} (npm) in the scatterplot, however, a few classes form interesting outliers with high wmc values but low npm values. These unusual characteristics  correspond with the \emph{Functional Decomposition} smell (highlighted as black dots in the scatterplot in Figure~\ref{fig:captions}, right). As the other metric values for this set of classes are shown accordingly in the parallel coordinates plot (Figure~\ref{fig:captions}, left), we also observe that one of the classes is associated with large file size (loc) and a high number of bugs.}

In the \emph{Lucene 2.0} project (Figure~\ref{fig:teaser}), we observe good quality for coupling and inheritance, whereas quality is lower for complexity and cohesion. We see an outlier (\emph{StandardTokenizerTokenManager}) in the complexity bar chart, and when clicking it (see Figure~\ref{fig:teaser}), we observe it highlighted also in the explanations of code smells and the bug history section. The class that has three code smells (\emph{Large Class}, \emph{Functional Decomposition}, and \emph{Spaghetti Code}) and has been associated with bugs. \emph{StandardTokenizerTokenManager} also turns out to be a prominent outlier in the parallel coordinates plot and scatterplot. 

We then switch to code smell analysis and notice that \emph{Functional Decomposition} is the most occurring code smell, while we only see single instances of \emph{Large Class} and \emph{Spaghetti Code}. We further investigate the classes that carry \emph{Functional Decomposition} by exploring the overview visualizations and source codes to discern the common reason for the code smell. We recognize that the majority of the classes that have \emph{Functional Decomposition} share a problem: more private methods and fewer public methods; moreover, they have large class sizes.

\section{Lessons Learned} \label{sec:lessons}
Finally, we summarize our experience gained in designing the system as lessons learned. We focus on aspects that generalize beyond the specific application in software engineering to interfaces producing interactive documents that consist of generated textual and visual data descriptions.

\emph{Overview always!} -- While, in traditional text documents, text and graphics are arranged in a fixed layout, we observed that loosening this tight coupling is beneficial for interactive documents like ours. Especially, overview visualizations should always stay in view and not\deleted{eventually} scroll out with the text (like tested in the first iteration of our formative evaluation). Also, texts that provide a high-level summary might act as anchors for the users to return to and should stay visible. We used the section describing code smells in this way. Such overview elements are important for interactive highlighting as well: hovering an element anywhere in the interface, related items in the overview representations get highlighted. For instance, with such highlighting in the code smell section, we can easily indicate by which code smells a highlighted class is affected. The always visible overview representations, together with a stable layout, provide a reliable skeleton for the user and allows other content to change dynamically without confusing the user's mental map. \added{This lesson relates to Shneiderman's information-seeking mantra~\cite{shneiderman2003eyes}, which emphasizes on overview first and details later.}

\emph{Consider brushing text, really.} -- From a visualization perspective, texts in an interface might appear being \emph{dead}---they are not part of a visualization and cannot be interacted with. However, we have shown that text can be integrated into interactions almost like any other element of a visualization. Brushing a marked text element, the related visual representations get highlighted across all visualizations, and vice versa. Opportunities where such interactions make sense appear naturally whenever describing a data-related entity (in our case: a class, a software metric, a code smell, etc.). Also, methodological and educational explanations as well as further data-driven details can be interactively linked with a text. For the text we generate, we tried to identify all elements the users might want to explore further and decided what would be an appropriate interactive linking. \added{This lesson links to the interaction model proposed by Beck and Weiskopf~\cite{beck2017word}.}

\emph{Captions! And make them dynamic.} -- Although not difficult to implement, many visual analytics systems even lack basic captions for the visualizations shown in different views. We decided to add captions to make the interface more self-explanatory. And it quickly turned out that these captions should to be adaptive: When something changes in the visualization based on an interaction, the caption needs to change accordingly as it should always accurately describe what is currently shown. Template-based natural language generation provides the means for implementing such adaptive captions.

\emph{Pointers, everywhere.} -- Although we defined them as separate categories in Section~\ref{sec:concepts}, data-driven, educational, and methodological explanations should somewhat blend in practice. A purely data-driven explanation might read cryptically, but with a few educational or methodological hints, the text would provide the necessary pointers to understand it more easily. Also, for instance, an educational explanation can profit from examples from the data, like we have integrated in the respective background text. These hints can be used as hyperlinks to the more detailed explanations. Still, we recommend, when authoring the texts, strictly discerning between the categories of text and also reflecting this categorization in the user interface by a consistent layout. For instance, we presented \emph{methodological explanations} only in tooltips of info icons and educational explanations, marked with the term \emph{background}, in the details view on the right. This consistency allows users to learn where to look for certain information.

\emph{You just learn on the side.} -- Like most visual analytics systems, our approach is also built to support users in understanding the specific data shown. With every data analysis, the users gain experience and might also have general insights with respect to the overall analysis procedure or domain. In contrast to other interfaces, we actively support this \emph{learning on the side} through methodological and educational explanations. Through activating these explanations, the interface is adapting---not automatically but with only little extra effort---to the individual information needs of the user.

\section{Conclusion and Future Work} \label{sec:conclusion}

We have introduced an approach that automatically generates interactive documents to describe multivariate data and tailored the approach in a design study for reporting the code quality of a software project. The approach is \emph{exploranative} as it \emph{explains} the data and background in a textual way as well as it supports the \emph{exploration} of the data through interactive visualizations. The textual and visual representations are consistently linked with each other.\deleted{ The interactions and visuals are designed to consistently and closely link the different representations and views.} \added{While our design process and formative evaluation have already provided insights into usability and aspects of visual design, a next step is to conduct a larger-scale user study with representatives of our targeted users (\deleted{i.e., stakeholders involved in software engineering,}see Section~\ref{sec:targeted_users}). A goal of the study will be to test in a realistic setting whether actionable results can be derived from our documents across participants with different levels of expertise in code quality analysis.}

\added{We have introduced the general concept of \emph{exploranative documents for multivariate data} (Section~\ref{sec:concepts}) and have demonstrated this concept in detail with one example from software engineering. Still, future work is to show the applicability of the concept for other data analysis scenarios. For instance, in the context of \emph{Industry 4.0}, comparable documents could summarize multivariate performance and maintenance indicators of the machines of a modular production line. Besides visualizing multivariate raw data, textual descriptions of analysis results could directly hint at potential issues before the issues manifest themselves in an expensive halt of the production line. A comparable interactive document could be the result of another design study process, then involving manufacturing researchers and professionals.}

We consider our contributions in extending interactive textual documents with respect to explorative analysis as a step toward blurring the borderline between textual and visual data representations. The interaction model introduced to connect visual and textual elements goes beyond previous work and showcases how text generation and brushing-and-linking techniques can play together in a multi-view system. We believe that texts, well-integrated with visualizations, can make data analysis more accessible and easier to understand for a wide audience. With further integration, eventually, \emph{text} and \emph{visualization} will become only two points in a continuum (with any point in between possible) instead of being treated as two separate modalities.

As part of future work, we are interested in investigating the educational aspects of visual analytics in more detail. For instance, learning research literature~\cite{wittwer2010effective} discusses \emph{worked examples}---step-by-step demonstrations---and \emph{instructional explanations} that both share similarities to the textual explanations used in our approach. Although learning is usually not the primary goal of a visual analytics system, it still would be highly relevant to study how such systems educate their users. Follow-up research questions are whether adding educational and methodological explanations to an interactive visualization has positive educational effects \added{and how users like and deal with textual descriptions when they have already become familiar with the text}.

\acknowledgments{Fabian Beck is indebted to the Baden-W\"urttemberg Stiftung for the financial support of this research project within the Postdoctoral Fellowship for Leading Early Career Researchers.}

\bibliographystyle{abbrv-doi}

\bibliography{references}
\end{document}